\documentclass[
  draft
]{agujournal2025}

\usepackage{amsmath}
\usepackage{float}
\usepackage{overpic}
\usepackage{subcaption}
\usepackage{pdfpages}
\justifying % XL XL XL

\begin{document}

\journalname{Journal of Advances in Modeling Earth Systems}

% Your title can be multiple lines long.
\title{Improved Precipitation Variability in the Community Atmosphere Model through Implementation of a Neural-Network Deep Convection Scheme}

\authors{%
                  Xavier J. Levine\affil{1}, %\thanks{their affiliation address and/or funding information},
                  Griffin Mooers\affil{2}, %\thanks{their affiliation address and/or funding information},
                  Judith Berner\affil{3}, %\thanks{their affiliation address and/or funding information},
                  Jack W. Atkinson\affil{4}, %\thanks{their affiliation address and/or funding information},
                  Marat F. Khairoutdinov\affil{5}, %\thanks{their affiliation address and/or funding information},
                  Janni Yuval\affil{6}, %\thanks{their affiliation address and/or funding information},
                  Dominic Orchard\affil{4,7}, %\thanks{their affiliation address and/or funding information},
                  Paul A. O'Gorman\affil{2} %\thanks{their affiliation address and/or funding information}
                  }

% Repeat for all authors:	      
\affiliation{1}{Department of Earth and Environmental Engineering, Columbia University, New York, NY, USA} % First argument is their footnote number, as above.
\affiliation{2}{Department of Earth, Atmospheric and Planetary Sciences, Massachusetts Institute of Technology, Cambridge, MA, USA} % First argument is their footnote number, as above.
\affiliation{3}{U.S. National Science Foundation National Center for Atmospheric Research, Boulder, CO, USA} % First argument is their footnote number, as above.
\affiliation{4}{Institute of Computing for Climate Science, University of Cambridge, Cambridge, UK} % First argument is their footnote number, as above.
\affiliation{5}{School of Marine and Atmospheric Sciences and Institute for Advanced Computational Science, Stony Brook University, Stony Brook, NY, USA} % First argument is their footnote number, as above.
\affiliation{6}{Google Research, Mountain View, CA, USA} % First argument is their footnote number, as above.
\affiliation{7}{School of Computing, University of Kent, UK} % First argument is their footnote number, as above.

% Corresponding author. Do not prepend with ``SI Corresponding author: '' in published version.
\authoraddr{%
                      % e.g., their full name, their department, their academic institution, their building, their City, State, Zip, Country. (theiremail\@ their institution)
                      Xavier J. Levine, Department of Earth and Environmental Sciences, Columbia University, 556-7 Schermerhorn Hall Extension, New York, NY, USA (xl3521@columbia.edu)
                   }

\authorrunninghead{Ignored.} % Ignored.

% Key points.  
%   Takes three arguments.  Remember to specifiy them all, using {} for no item.
%
\keypoints%
% {}{}{}
%    % First key point, etc.
    {A neural network trained on a storm-resolving simulation replaces the default deep convection scheme in CAM6 to give CAM-NN}
    {CAM-NN runs stably for 15 years with modest mean biases despite no retuning of other subgrid parameterizations}
    {CAM-NN has substantially improved precipitation intensity, precipitation extremes, and the timing of the diurnal cycle over land}    
%    {Summarize the main points and conclusions of the article}
%    {Each must be 140 characters or fewer with no special characters or punctuation and must be complete sentences}

% Call after \authors.

% Title running heads will be ignored, too:
\titlerunninghead{Ignored.}

% Key points was here

\maketitle

\bigskip  % Remove; for demo purposes only.
% OR:

\begin{abstract}
    Conventional deep convection schemes remain a primary source of uncertainty in global climate models, yet replacing them with data-driven alternatives while maintaining physical consistency and numerical stability is an open challenge. We present a hybrid configuration of the Community Atmosphere Model version 6 (CAM6), in which a fully connected deep neural network replaces the Zhang–McFarlane deep convection scheme (ZM), while radiation, turbulence, and other cloud processes are left unchanged. The network is trained on output from a storm-resolving global model - the global System for Atmospheric Modeling (gSAM) -  using a coarse-graining approach to derive subgrid-scale convective fluxes and tendencies. This data-driven parameterization is ported to CAM6 and found to run stably for more than a decade while conserving energy and moisture. It produces a broadly realistic climatology without retuning the remaining parameterizations. Relative to the default Zhang-McFarlane scheme, the neural network parameterization substantially improves some long-standing model deficiencies. In particular, the biases in light precipitation and extreme precipitation are markedly reduced, and the diurnal cycle of tropical land precipitation shifts from a spurious midday peak over many regions to the observed late-afternoon peak. These results support targeted ML-based convection schemes as a viable path toward reducing systematic model errors in next-generation climate models.
     \begin{plainlanguagesummary}
           Numerical models are essential for understanding how Earth's climate works and how it will change, but they struggle to represent processes smaller than their grid spacing, like the deep, rain-bearing clouds that carry heat and moisture upward in the tropics. Climate models represent these clouds with simplified physical approximations, which are a major source of error in properties that matter most to society, especially the intensity and timing of rainfall. Here, we replace one such approximation with a machine learning (ML) model trained on a highly detailed simulation that explicitly resolves individual storms but which is too expensive to run for long. We embed the ML model inside a climate model in place of the original approximation. The resulting hybrid model runs stably for over a decade and substantially improves long-standing problems found in the default model, including unrealistic rainfall amounts and timing, showing that ML can meaningfully improve how climate models represent clouds.
     \end{plainlanguagesummary}
     %
     % Do not put more abstract text body here!
\end{abstract}

\section{Introduction}\label{sec:intro}  % Or first section name if you have not introduction.

Convection schemes are a critical component of global climate models because of their role in transporting heat and moisture vertically, representing phase transitions of water, and generating precipitation \cite{arakawa2004cumulus}, and they have been recognized as a source of uncertainty in climate projections, primarily through their impact on tropical convection-circulation coupling \cite{bony2015clouds} and cloud-radiative feedbacks \cite{webb2015impact}. Current convection schemes are usually based on idealized models for cloud structure together with physics-based ideas for what drives convective instability and how convection interacts with and influences the large-scale environment (e.g., \citeNP{kuo1974further,arakawa1974interaction,emanuel1991scheme,betts1993betts,zhang1995sensitivity,tan2018extended}). Convection schemes strongly influence key characteristics of variability, including the diurnal cycle \cite{hohenegger2013controls}, precipitation intensity and extremes \cite{wilcox2007frequency}, and the Madden-Julian Oscillation and El Niño–Southern Oscillation \cite{wu2007impact}. The complexity of representing subgrid-scale convection and its embedded multiphase microphysical processes makes further theoretical progress challenging, which explains why most convection schemes are based on decades-old models and concepts. Nonetheless, there has been progress including development of more unified approaches \cite{arakawa2013unified, wu2014unified,tan2018extended} and super-parameterizations \cite{grabowski1999crcp,khairoutdinov2005simulations}.

Advances in machine learning (ML), combined with substantial increases in computational power and the growing availability of observational datasets and high-resolution simulations, have enabled the development of data-driven convection schemes. These data-driven schemes offer a potentially transformative pathway for improving the representation of convection in climate models (e.g., \citeNP{gentine2018could,rasp2018deep,ogorman2018using}). Hybrid modeling approaches, which retain the physically based dynamical core while replacing selected parameterizations (e.g., convection) with data-driven models, have the potential for greater flexibility, generalizability and interpretability compared to fully end-to-end ML frameworks. In these hybrid configurations, large-scale dynamics governed by fundamental physical laws are preserved, while machine learning is used to represent unresolved processes such as convection and cloud microphysics (e.g., \citeNP{rasp2018deep,yuval2020stable,kochkov2024neural}). This framework reflects a natural separation between model components that represent well-understood physics at resolved scales and those components that are either inherently empirical or physically understood but unresolved at the model grid scale; from this perspective convection and cloud microphysics are well suited for ML-based parameterization.

Previous work has explored a range of ML approaches for representing moist processes in climate models, spanning from unified to more targeted techniques. In a unified framework, a single ML model is used to represent both subgrid-scale and grid-scale moist physics and in some cases boundary layer turbulence and radiative processes (e.g., \citeNP{rasp2018deep,kochkov2024neural,han2025decadal,wattmeyer2024neural}). A more targeted approach seeks to improve the representation of a specific process, such as warm rain microphysics \cite{gettelman2021machine}. In an aquaplanet setting, \citeNP{yuval2020stable} and \citeNP{yuval2021use} included multiple processes such as moist physics, turbulence and radiation but predicted the different processes separately to improve physical consistency and conservation of energy and water.

Despite these advances, the implementation of a data-driven scheme trained on a high-resolution simulation to represent deep convection within the standard physical decomposition of a global climate model (GCM) remains untested. This is challenging for at least two reasons: (1) finding training data that can isolate deep convection from other subgrid- or grid-scale mixing processes is inherently difficult, and (2) implementing such a scheme may require retuning the other model components, since these are typically tuned based on the existing convection scheme. Recalibrating every other parameterization to accommodate a new convection scheme is itself a substantial undertaking, and an alternative path could be to fine-tune the neural-network convection scheme itself online, holding the other parameterizations fixed, using methods such as ensemble Kalman inversion \cite{christopoulos2024online}.

Here, we adopt a targeted approach in which a fully-connected deep neural network is used to represent deep convection in the CAM6 atmospheric model which is part of the Community Earth System Model version 2 (CESM2) \cite{danabasoglu2020community}. The neural-network parameterization replaces the Zhang-McFarlane deep convection scheme \cite{zhang1995sensitivity}. We do not replace physical parameterizations for other subgrid processes (e.g., radiation, non-convective cloud processes and turbulence). By focusing on deep convection and excluding non-convective cloud processes from the parameterization, we lessen the extent to which our parameterization must perform well for low clouds or in regions such as high latitudes which can be difficult for hybrid approaches \cite{wattmeyer2024neural,han2026exploring}. Our rationale for not including radiation in the parameterization is that the resulting hybrid climate model could then be generalizable across a wider range of climate scenarios and external forcings, including, for example, different aerosol and ozone distributions that could be difficult to include in training datasets from an expensive global convection resolving model.

Our parameterization is trained on a convection-resolving simulation using the global System for Atmospheric Modeling (gSAM) \cite{khairoutdinov2022global} from the DYnamics of the Atmospheric general circulation Modeled On Non-hydrostatic Domains (DYAMOND) Winter Intercomparison Project (DYAMOND2) which is evaluated by \citeNP{in2026evaluation} and builds on the first summer DYAMOND phase \cite{stevens2019dyamond}. At storm-resolving resolution, gSAM explicitly simulates convection, mesoscale convective systems, and cloud-circulation interactions that remain poorly represented in coarse GCMs such as CAM6, making it a physically consistent and information-rich training dataset. We apply a coarse-graining approach to calculate subgrid vertical fluxes and sources/sinks in a column following \citeNP{yuval2020stable} and \citeNP{yuval2021use}. This design aims to enhance physical consistency and improve generalizability across climates, while maintaining compatibility with existing GCM architectures. Because gSAM has a different vertical grid and prognostic variables compared to CAM6 - differing not only in vertical resolution but also in vertical coordinate (height-based versus hybrid sigma-pressure), and in underlying dynamical assumptions (anelastic versus hydrostatic) - we also test the extent to which a data-driven convection scheme may be ported to substantially different target atmospheric models and different vertical resolutions.

Our results demonstrate that a hybrid earth system modeling framework can successfully incorporate a data-driven representation of atmospheric convection while maintaining numerical stability and adherence to fundamental physical constraints, including conservation of energy and water. The resulting CAM6 model with the neural network convection scheme not only preserves a realistic mean climate but also improves upon the default CAM6 configuration across several key diagnostics. Most notably, the implementation of the neural network convection scheme substantially enhances the representation of key aspects of precipitation variability in CAM6 – namely the diurnal cycle and intensity distribution of precipitation – both of which are a long-standing challenge for global climate models, highlighting the potential of targeted ML-based parameterizations to advance next-generation climate modeling.

In Section~\ref{sec:model_description}, we describe how we trained our deep neural network convection scheme using output from the storm-resolving model gSAM, and how we implemented it in the Earth System Model, CESM2-CAM6. We present a comparison of CAM6 simulations with the neural-network convection scheme as compared to observations and simulations with default CAM6 in Section~\ref{sec:results} before giving conclusions in Section~\ref{sec:conclusions}.

\section{Models, training and implementation}\label{sec:model_description}

\subsection{Storm-Resolving Model and Simulation used for Training the Neural Network Convection Scheme: gSAM}\label{subsec:gSAM}

We leverage hourly output from a high-resolution global simulation with gSAM \cite{khairoutdinov2022global} as training data for the neural network parameterization. The gSAM simulation was run as part of the DYAMOND2 project and encompasses 40 days of boreal winter (January 20\textsuperscript{th}, 2020 to March 1\textsuperscript{st}, 2020) \cite{klocke2019project}. DYAMOND also includes an earlier phase, DYAMOND1 \cite{stevens2019dyamond}, covering 40 days of boreal summer (August 1\textsuperscript{st} to September 10\textsuperscript{th}, 2016); however, the coarse-grained output needed to train the neural network was available only for DYAMOND2. A longer time period and seasonal range for training would have been preferable, but we nonetheless expect that the global coverage at storm-resolving scales helps offset the modest temporal range of our training data, with spatial diversity aiding the model's ability to generalize across temporal scales. The horizontal grid spacing of the DYAMOND2 gSAM simulation varies with latitude and is in the range 2-4km except at high latitudes, and there are 74 vertical levels. 

gSAM is a global version of the System for Atmospheric Modeling (SAM), a mesoscale model on a Cartesian grid \cite{khairoutdinov2003cloud}. gSAM represents flow over topography and has a land model to represent land-atmosphere interactions \cite{khairoutdinov2022global}. Unlike many other convection-permitting global atmospheric models, gSAM is an anelastic model on a latitude-longitude grid with height as a vertical coordinate \cite{khairoutdinov2022global}. The anelastic formulation makes the model efficient and more straightforward for calculation of subgrid terms, and this formulation may also explain why gSAM has different vertical velocity statistics compared to other global storm-resolving models \cite{mooers2023comparing}. Run globally at high resolution, gSAM represents deep convection without the need for convective parameterizations, making its output a rich testbed to generate neural-network convective parameterizations. 

\subsection{Climate Model for Hybrid Implementation of the Neural Network Convection Scheme: CAM6}\label{subsec:CAM6}

We implement our parameterization in the Community Atmosphere Model version 6 (CAM6) which is the atmospheric component of the Community Earth System Model version 2.1.5 (CESM2), a state-of-the-art earth system model \cite{danabasoglu2020community}. Here, CAM6 is run in an atmosphere-only configuration with prescribed historical sea surface temperatures (SSTs) and sea ice concentrations (SICs), along with time-varying anthropogenic and natural greenhouse gas concentrations and aerosol emissions following CMIP protocols.

In its standard configuration, CAM6 is built around a dynamical core that solves the primitive equations of the atmosphere on a latitude–longitude grid using a conservative, flux-form formulation. This core employs a finite-volume discretization that ensures the conservation of mass, energy, and tracers. CAM6 includes state-of-the-art physics-based representations of cloud microphysics, radiation, subgrid-scale convection, and shallow and boundary layer turbulent mixing.

In the default version of CAM6, deep convection is parameterized using the Zhang-McFarlane (ZM) scheme \cite{zhang1995sensitivity}. The ZM scheme is a mass-flux parameterization that represents deep convection as an ensemble of entraining convective plumes rising from the boundary layer and interacting with the large-scale environment. Its closure assumption links convective activity to the consumption of dilute convective available potential energy (CAPE) over a specified relaxation timescale. Convection is triggered when CAPE exceeds a threshold, and the scheme then acts to reduce thermodynamic disequilibrium between plumes and the environment. The longevity of ZM in CAM reflects its physically motivated yet computationally efficient design, which produces a reasonable representation of tropical precipitation and large-scale circulation \cite{neale2013mean,bogenschutz2018path}, although important deficiencies remain as shown below.

Cloud microphysical processes are represented by the Morrison-Gettelman scheme MG2, a two-moment parameterization that predicts both mass and number concentrations of cloud liquid, cloud ice, rain and snow \cite{gettelman2015advanced}.  Both boundary layer turbulence and shallow convection are represented by the Cloud-Layers Unified by Binormals (CLUBB) scheme \cite{bogenschutz2013higher}. Radiation is represented by a broadband radiative transfer scheme that accounts for the absorption, emission, and scattering of both shortwave (solar) and longwave (terrestrial) radiation by atmospheric gases, clouds, and aerosols \cite{iacono2008radiative}. Initial development of our parameterization implementation in CAM6 was greatly facilitated by the single column version of CAM6 \cite{gettelman2019single}.

\subsection{Structure of the parameterization}\label{subsec:NNCONV_description}

\begin{figure}[H]
\centering
\includegraphics[width=\textwidth]{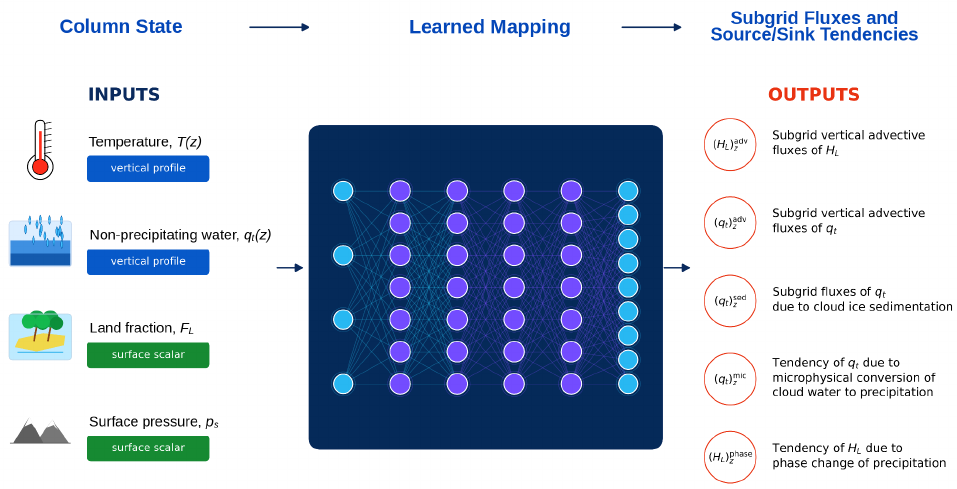} 
\caption{Structure of the neural-network parameterization, listing its inputs and outputs.}
\label{fig:Fig1}
\end{figure}

To operationalize our machine learning based parameterization, we embed dense, feed-forward neural networks in the CAM6 model. Much of the initial structure of the parameterization was derived from the aquaplanet parameterization of \citeNP{yuval2021use}, but the parameterization was significantly updated to match the gSAM simulation training data (which includes land and topography) and the resolution of the target CAM6 model and will hereafter be referred to as NN-CONV. An overview of the parameterization is given in Figure~\ref{fig:Fig1}.

We focus on thermodynamic and moisture variables and use a local column based structure for the neural network parameterization, consistent with the structure of existing physical parameterizations in CAM6. In the future, our approach could be extended to include momentum \cite{yuval2023neural} and non-local parameterizations \cite{wang2022non}. The column based structure for the neural network parameterization uses vertical profiles of the thermodynamic and moisture variables to prediction a set of energetic and cloud microphysical tendencies and vertical fluxes. The training is done using gSAM equations and variables with the conversion to CAM6 variables performed in the CAM6 implementation (see Section~\ref{subsec:NNCONV_implementation}). The prognostic gSAM variables that we use in the parameterization are the liquid-ice static energy ($H_L$) and total water mixing ratio ($q_T$), both computed using only non-precipitating condensate (i.e., cloud liquid and cloud ice, excluding rain, snow, and graupel):
\begin{subequations}
\begin{align}
&H_L = c_p T + g z - \left( L_c q_c + L_s q_i \right) \label{eq:HL}\\
&q_T = q_v + q_c + q_i \label{eq:QT}
\end{align}
\end{subequations}
Here, $c_p$ is the heat capacity of dry air, $T$ is temperature, $g$ is the gravitational acceleration, $z$ is the geopotential height, $L_s$ and $L_c$ are the latent heat of sublimation and condensation, respectively, and $q_v$, $q_c$, and $q_i$ are mixing ratios for the vapor, liquid and ice phase of non-precipitating water. 

Each column prediction is derived from vertical profiles $T$ and $q_T$, augmented with surface reference pressure $p_s$ and land fraction ($F_L$) as additional scalar inputs. Extending Yuval et al. (2021) by including $p_s$ and $F_L$ allows the network to distinguish between different surface elevations and between land and ocean grid points. The outputs of NN-CONV are 5 column vectors: the subgrid-scale vertical advective fluxes of $H_L$ and $q_T$, the microphysical conversion of water between non-precipitating and precipitating components, the sedimentation mass flux of ice condensate, and the thermodynamic effect of phase changes in precipitating water. These outputs define a set of energetic and microphysical tendencies and vertical fluxes that affect the liquid-ice static energy ($H_L$) and total water mixing ratio ($q_T$) budget at a given vertical level of the atmosphere:
\begin{subequations}
\begin{align}
&\partial_t (H_L)^\mathit{NN} = -(1/\rho_0) \, \partial_z (H_L)^{adv} - L_p (q_T)^{mic} + L_f (1/\rho_0) \, \partial_z (q_T)^{sed} + (H_L)^{phase} \label{eq:HL_NN} \\
&\partial_t (q_T)^\mathit{NN} = -(1/\rho_0) \, \partial_z (q_T)^{adv} + (q_T)^{mic} - (1/\rho_0) \, \partial_z (q_T)^{sed} \label{eq:qT_NN}
\end{align}
\end{subequations}
where $\rho_0(z)$ is the anelastic reference density profile in gSAM, and $L_p = L_c + L_f (1-\omega_p)$ is an effective latent heat for precipitating water where $L_f$ is the latent heat of fusion and $\omega_p$ is a partition function corresponding to the fraction of ice in precipitating water (dependent solely on temperature). $(H_L)^{adv}$ and $(q_T)^{adv}$ are subgrid-scale vertical advective fluxes of $H_L$ and $q_T$, respectively, calculated as
\begin{subequations}
\begin{align}
&(H_L)^{adv} = \rho_0 \, \overline{w' H_L'} \label{eq:HL_adv} \\
&(q_T)^{adv} = \rho_0 \, \overline{w' q_T'} \label{eq:qT_adv}
\end{align}
\end{subequations}
where $w$ is the vertical wind, the overline denotes a spatial coarse graining, and the primes denote subgrid variations from the coarse-grained mean (see Section~\ref{subsec:NNCONV_training}). $(q_T)^{mic}$ is the microphysical conversion of water between non-precipitating and precipitating components. $(q_T)^{sed}$ is the sedimentation mass flux of ice condensate (negative values for downward flux). $(H_L)^{phase}$ is the thermodynamic effect of phase changes in precipitating water (e.g., graupel melting to rain) calculated as:
\begin{equation}
(H_L)^{phase} = (1/\rho_0) \, (\partial_z L_p) \left( \rho_0 \, \overline{w' q_p'} + F_{qpz} - P_{tot} \right) \label{eq:HL_phase}
\end{equation}
where $q_p$ is the precipitating water mixing ratio, $F_{qpz}$ the vertical diﬀusive flux of precipitating water, and $P_{tot}$ the precipitation flux which is positive downwards.

The parameterization does not include precipitating water as a prognostic variable, so we diagnose the convective precipitation from NN-CONV ($P^\mathit{NN}$) as the column-integrated loss of $q_T$ that results from the tendencies from NN-CONV in Equation \eqref{eq:qT_NN}:
\begin{equation}
P^\mathit{NN} = - \int_{z_s}^{\infty} \rho_0 \, \partial_t (q_T)^\mathit{NN} \, dz = -\int_{z_s}^{\infty} \rho_0 \, (q_T)^{mic} \, dz - (q_T)^{sed}\Big|_{z_s} \label{eq:Pc}
\end{equation}
where $z_s$ is the height at the surface.

The 5 outputs of NN-CONV $\left( (H_L)^{adv},\ (q_T)^{adv},\ (q_T)^{mic},\ (q_T)^{sed},\ (H_L)^{phase} \right)$ align with the thermodynamic and moisture budget of our training model, gSAM. As in \citeNP{yuval2021use}, NN-CONV predicts vertical fluxes for advection and sedimentation rather than tendencies directly. Because the subgrid advective tendency at each level is computed as the divergence of these fluxes, setting the fluxes to zero at the top and bottom of the atmosphere ensures column-wise conservation of $H_L$ and $q_T$. We note that the microphysical tendencies $(q_T)^{mic}$ and $(H_L)^{phase}$ aggregate a large number of processes but do not include radiative heating.

Overall, NN-CONV has 100 inputs comprising two 49-level column vectors for temperature and non-precipitating water plus the two surface scalars and 245 outputs made up of the five 49 level columns of energy and moisture tendencies or fluxes (Figure~\ref{fig:Fig1}). The vertically resolved variables are interpolated to a sigma coordinate as discussed in detail in the next section. 

We find that a densely connected feed-forward architecture with five hidden layers, when trained on sufficiently large datasets, achieves high skill in offline statistics without requiring more complex architectures such as convolutional layers or long short-term memory networks, consistent with a previous study using neural networks trained on simulation output with realistic geography from Superparameterized CAM (SPCAM) \cite{mooers2021assessing}. This architecture also facilitates later coupling to CAM6, where the candidate neural network parameterization can be implemented in Fortran with relative ease. This does not imply that hyperparameter choices are unimportant: we rigorously tuned the loss function, learning rate, and the number and size of hidden layers, among other parameters (see Section~\ref{subsec:NNCONV_training}). Empirically, however, we find that at CAM6’s horizontal resolution, once the hyperparameters are broadly appropriate for the column-to-column prediction task, offline performance is more sensitive to training data volume than to specific architectural choices.

\subsection{Horizontal Coarse-Graining and Vertical Interpolation of gSAM Output and Training of the Neural Network}\label{subsec:NNCONV_training}

The ultimate performance of NN-CONV once embedded in CAM6 is sensitive to how the data used to train the neural network parameterization is prepared. Coarse-graining was applied to both the neural network input variables $\left( T,\ q_T,\ p_s,\ F_L \right)$ and to calculate the output variables $\left( (H_L)^{adv},\ (q_T)^{adv},\ (q_T)^{mic},\ (q_T)^{sed},\ (H_L)^{phase} \right)$. We define the subgrid vertical advective flux of variable $x$ (i.e., $x = q_T$, $H_L$) as the coarse-grained vertical flux $\left( \rho_0 \, \overline{wx} \right)$ minus the flux due to the coarse-grained vertical velocity $\left( \rho_0 \, \overline{w}\,\overline{x} \right)$, i.e. $\rho_0 \, \overline{w'x'} = \rho_0 \, \overline{wx} - \rho_0 \, \overline{w}\,\overline{x}$. This subgrid-scale decomposition is applied to the vertical advective fluxes of $H_L$ and $q_T$, and the corresponding subgrid advective fluxes $(H_L)^{adv}$ and $(q_T)^{adv}$ are then predicted by the neural network. The resolved advection is assumed to be handled by the advection scheme in CAM6. The moist physics outputs, namely the sedimentation flux $(q_T)^{sed}$, microphysical conversion $(q_T)^{mic}$, and heating from phase change of precipitation $(H_L)^{phase}$ are the total quantities rather than subgrid-scale contributions. We use the total moist physics outputs because it is not possible to cleanly disentangle the subgrid components from $(q_T)^{mic}$ and $(H_L)^{phase}$ because they depend on precipitating water $q_p$ which is not a prognostic variable in the parameterization. This could lead to some double counting of microphysical processes by NN-CONV and the MG2 scheme when and where NN-CONV is triggered, although that is expected to be reduced by the updates from NN-CONV being applied prior to calling MG2. Calculating an approximate subgrid decomposition of the microphysical outputs for  training NN-CONV could be the subject of future work.

As a first step, we calculated and stored the required spatially coarse grained fields (including variables, fluxes and tendencies) with a coarse-graining factor of 12 as the DYAMOND2 gSAM simulation was run. Calculating coarse-grained fluxes and tendencies as the simulation was run improved accuracy and avoided the need for expensive post-processing calculations at the original high-resolution. Because this intermediate grid still differed from our target model, we then needed to map the fields to the exact horizontal grid of CAM6 $(\sim 100 \text{ km})$. To accomplish this, we applied conservative regridding via the xESMF package in Python (Zhuang et al., 2025). This regridding method (and the initial coarse graining) preserve integrated quantities, ensuring the integrated fluxes and tendencies over an area of the globe remain consistent after moving from the gSAM grid to the CAM6 grid. Coarse graining to the exact CAM6 grid ensures that the neural network parameterization is trained at the appropriate resolution to predict the correct subgrid terms it will need when coupled in CAM6.

Beyond horizontal resolution differences in the initial and target model data, the realistic land component of gSAM presents another challenge in preprocessing. Because gSAM output is defined on a fixed-height (geopotential) grid, many vertical levels over elevated terrain fall below the surface, leaving those points undefined; training a neural network directly on such data is therefore problematic. To better learn the effects of topography while avoiding this issue, NN-CONV requires the atmospheric columns to be interpolated from their native height coordinates to sigma levels $\sigma$, where all values remain defined above the surface even over topography. Here, the coordinate axis $\sigma$ is defined as pressure normalized by the surface pressure (where pressure refers to the anelastic reference pressure profile in gSAM). We use a simple linear interpolation to map input and output variables from the native height grid of the gSAM model to a sigma coordinate grid, where the sigma coordinate grid is set equal to the sigma values obtained from the anelastic reference pressure grid over ocean. The inclusion of surface pressure as an input to the neural network means that there should not be loss of information from interpolating to sigma over topography. Following Yuval et al 2021, we choose a cutoff level near the tropopause above which the inputs and outputs are not used for the neural network, with the cutoff here chosen to be $\sigma = 0.15$, leaving 49 sigma levels in total.

After coarse graining and interpolation, the gSAM simulation output is at the resolution $(\sim 100 \text{ km})$ needed to match the CAM grid, and in sigma coordinates, which avoids undefined values below topography.

A final preprocessing step normalizes the neural network inputs and outputs. For each variable, we subtract the mean and divide by the standard deviation. For vertically resolved inputs, the normalization is applied separately at each vertical level. This distinction is especially important for the non-precipitating water mixing ratio $q_T$ input, whose magnitude varies strongly with height. By contrast, the output variables are tendencies or fluxes whose vertical profiles do not typically span orders of magnitude between the lower and upper troposphere. Column-wide normalization is therefore appropriate because it preserves the physically meaningful differences in magnitude across levels while still placing each output variable on a comparable numerical scale for training.

With this preprocessed data, we can now deploy the bulk of this global simulation (34 days and 132 GB of data), randomized with respect to time, to train the neural network itself. The weights of NN-CONV are obtained by minimizing a Huber loss function applied to all output of the MLP, using backpropagation (see Table~\ref{tab:Table1} for details). A portion of the remainder of the global simulation data, 4 days, is held out to test the trained neural network parameterization offline for accuracy of predictions. When testing offline, we found the most promising candidate neural network parameterizations are produced by keeping the learning rate relatively low (0.002) and training over a large number of epochs (55). Additional skill was extracted by adding a 10x learning rate decay late in the training (after 40 epochs), as well as using a uniform hidden-layer width of 128 nodes (Table~\ref{tab:Table1}). Total training time when run on one CPU node (with 32 tasks) in the MIT Engaging cluster takes approximately 77 hours (CPUs were used rather than GPUs due to the lengthy training time and the particular wall clock limits of the GPUs on this cluster). Once ready, the weights are saved to a NetCDF file for later embedding into CAM6.

The most promising candidate neural network parameterizations is able to closely predict the mean and variance of the fluxes over the entire troposphere over both land and ocean (Figure~S1 in Supporting Information S1). This good performance is an indicator that the neural network parameterization is ready for coupling into the CAM6 model. 

\subsection{Implementation of the Neural Network Parameterization in CAM6}\label{subsec:NNCONV_implementation}

The NN-CONV scheme replaces the \citeNP{zhang1995sensitivity} deep convection parameterization in CAM6, which represents deep convection using a plume model. Specifically, NN-CONV replaces the ZM-predicted convective tendencies in temperature, specific humidity, and mass fractions of non-precipitating liquid and ice condensate, as well as the related convective precipitation and number tendencies for cloud ice and liquid. NN-CONV is implemented in CAM6 via Fortran modules that define the fully connected neural network and the functions required to transform and interpolate CAM6 inputs to the neural networks vertical grid and variables, and the neural-network outputs back to the CAM6 vertical grid and variables. NN-CONV uses the weights that have been trained offline on the gSAM data, which are read in from a NetCDF file and used by the neural network scheme to produce the required outputs at every physics timestep.

In converting between gSAM and CAM6, we take advantage of the isomorphism described in section~\ref{subsec:NNCONV_description} of \citeNP{khairoutdinov2022global} between the anelastic equations in height coordinates (used in gSAM) and the primitive equations in pressure coordinates (used in CAM6). According to this isomorphism, variables are remapped in the vertical by equating the reference pressure profile in gSAM with the pressure profile in CAM6. In practice, the neural network outputs are in sigma coordinates, and so we use the sigma coordinate rather than pressure for the remapping.

Our convection scheme primarily outputs tendencies of temperature, specific humidity, and mass fractions of liquid and ice non-precipitating condensate. Convective precipitation is directly diagnosed from the net column loss in non-precipitating water (due to microphysical conversion to precipitation and any surface sedimentation of ice; see Equation \eqref{eq:Pc}) from the overall CAM6-converted tendencies. Number concentration tendencies for cloud liquid and ice are also diagnosed as described below.

\begin{figure}[H]
\centering
\includegraphics[width=\textwidth]{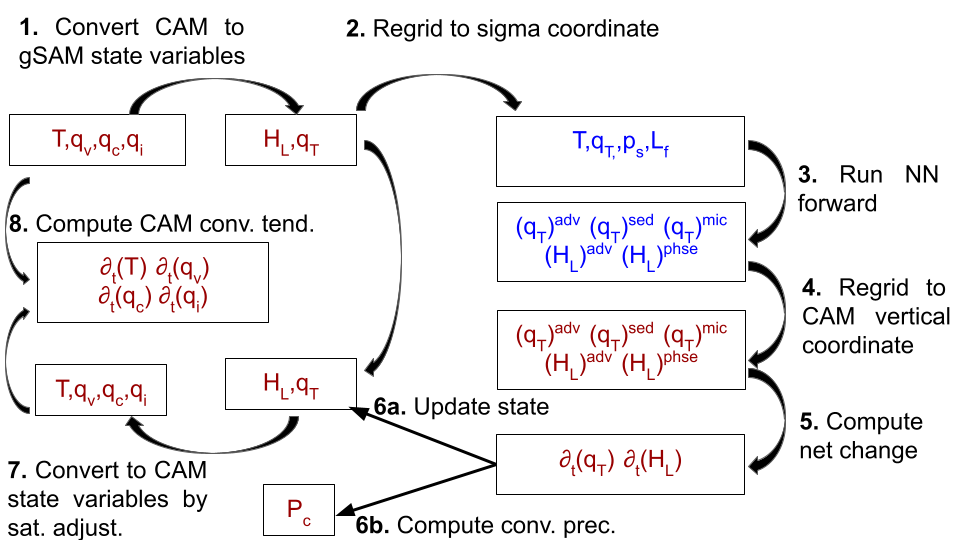}
\caption{Schematic of the implementation of the neural network parameterization in CAM6. Red text indicates steps carried out on CAM's native vertical grid; blue text indicates steps carried out on the neural network's sigma-level coordinates.}
\label{fig:Fig2}
\end{figure}

The parameterization proceeds in the following steps (Fig.~\ref{fig:Fig2}):

1) CAM temperature $T$ and moisture variables (specific humidity $q_v$ and mass fractions for cloud liquid $q_c$ and ice $q_i$) are converted to the gSAM energy and moisture variables ($H_L$, $q_T$) introduced in section~\ref{subsec:NNCONV_description}.

2) Vertical profiles of $T$, $H_L$ and $q_T$ are linearly interpolated from the pressure coordinate in the physics subroutines of CAM6 to the sigma coordinate used in training the neural network.

3) The vertical profiles of $T$ and $q_T$ on sigma levels together with surface pressure and land fraction are fed into the NN-CONV module, which returns its five outputs on sigma levels.

4) The five outputs are then remapped using conservative remapping to the pressure coordinate in CAM6.

5) Tendencies of $H_L$ and $q_T$ are calculated from equations \eqref{eq:HL_NN} and \eqref{eq:qT_NN} except that the vertical derivatives are calculated in pressure coordinates according to $-(1/\rho_0) \, (\partial X/ \partial z) = g \, (\partial X/\partial p)$ for each relevant vertical flux $X$.

6a) The surface precipitation rate is calculated using equation \eqref{eq:Pc} except the vertical integral is done in pressure rather than $z$, using the hydrostatic relation $\rho_0 \, dz = -dp / g$.

6b) The profiles of $H_L$ and $q_T$ are updated using the calculated tendencies and the CAM6 physics timestep.

7) The updated $H_L$ and $q_T$ are converted back to CAM6 variables ($T$, $q_v$, $q_c$ , $q_i$) using the moist thermodynamics of gSAM which assumes a partitioning of non-precipitating water $q_T$ into vapor, cloud, and ice dependent on temperature.

8) Lastly, tendencies of the CAM6 variables are calculated based on their initial and final states and the CAM6 physics timestep, and the tendencies are passed back into the main physics routine of CAM6.

The timestep used in gSAM (10s) is much shorter than the CAM6 physics timestep (30 minutes), but we find in practice it is not necessary to subcycle the neural network parameterization, presumably because the spatial coarse-graining increases the effective timescale of the parameterized physics. During the implementation, we enforce strict non-negativity in moisture by applying a local clipping correction: once all tendencies predicted by NN-CONV have been added to $q_T$ and $H_L$ in step 6b, any grid point where $q_T$ is negative is reset to a small positive floor value $(10^{-10} \text{ kg/kg})$. To maintain energetic consistency, an equivalent amount of energy is removed from $H_L$ at the same grid point following Equation \eqref{eq:HL}.

In order for the neural-network parameterization to function as a deep convection scheme, we turn it on only whenever deep convection is predicted to occur. We do this using a convective trigger that represents the need for instability to build up in the air column before it can be released by convective mixing. Following CAM6's default deep convection scheme ZM, we define a minimum value for the entraining convective available potential energy (CAPE) that needs to be reached before deep convection can occur. As in the ZM scheme, we set this threshold value at $70 \text{ J kg}^{-1}$. CAPE is computed from the buoyancy of an air parcel lifted from the surface that entrains environmental (grid-scale) air as in the ZM scheme.

NN-CONV is also deactivated whenever the predicted convective precipitation is zero or negative, ensuring that deep convection is always accompanied by precipitation. This distinguishes deep convection from non-precipitating shallow convection which is handled separately by CAM6's unified shallow convection and subgrid turbulence scheme, CLUBB.

Beyond tendencies of temperature, specific humidity, and condensate mass fractions, CAM6 requires the convection scheme to provide some additional outputs. The most important are tendencies for liquid and ice condensate number concentrations, which are needed for accurate representation of condensate in the microphysics and radiation scheme. Whenever the mass fraction tendency is positive, we derive number tendencies from the mass fraction tendency via a mass-to-volume conversion that assumes a uniform bulk ice particle mass, following the approach used in the ZM scheme. However, unlike ZM which always produces positive cloud liquid and ice tendencies through detrainment, NN-CONV can produce either positive or negative cloud tendencies. In the case of negative tendencies, we found it is more robust to assume the number concentration tendency is given by the existing number concentration times the mass fraction tendency and divided by the existing mass fraction. The atmospheric state is sensitive to the choice of mass-to-volume conversion factor for positive tendencies, particularly for ice condensate: too large of a conversion factor can produce large changes in the atmospheric state, especially in the tropical upper troposphere. We therefore treat the ice condensate mass-to-volume conversion factor as a tuning parameter for NN-CONV in CAM6. Based on the performance of the NN-CAM simulations in matching observations, we use a bulk ice particle mass of 654 nanograms, compared to the default CAM6 value of 327 nanograms. 

CAM6 also requires a deep convective mass flux which it uses to calculate the contribution of deep convection to cloud fraction. We compute this mass flux using the ZM scheme formulation; its effect on the atmospheric state is minor. In practice, in order to calculate both the ZM convective mass flux and CAPE trigger, we simply call ZM but prevent it from altering the atmospheric state. A more computationally efficient approach could be used in future work. Note that for simplicity we do not implement any convective momentum tendencies in NN-CAM.

While NN-CONV is implemented in CAM6, its design and training are largely independent of CAM6, and as such it could be implemented in other GCMs after retraining the neural network on the appropriate horizontal grid and with proper calibration once implemented.

\subsection{CAM6 Simulations}\label{subsec:CAM6-NNCONV}

We perform simulations with CAM6 using the standard ZM convection scheme (CAM-ZM) and with our neural network–based convection scheme (CAM-NN). No retuning of model parameters outside of the convection scheme was performed following the implementation of the NN scheme. While this approach enables a direct assessment of the impact of the NN scheme on CAM6’s climatology relative to the default configuration, it may lead to degraded performance of CAM-NN when evaluated against observations.

CAM6 is run with the CFMIP Observation Simulator Package (COSP), which enables direct comparison of simulated cloud properties with satellite observations \cite{swales2018cfmip}. COSP applies instrument simulators to model output to generate observation-equivalent diagnostics (e.g., cloud fraction, optical depth, and cloud top properties), ensuring consistency between model fields and remote sensing retrievals.

Simulations are produced for the period spanning from January 1, 2000 to December 31, 2014, with only the final 14 years (01/2001–12/2014) retained for analysis. CAM-NN does not noticeably run slower than CAM-ZM ($\sim 3.0 \text{ hours}$ on 512 CPUs per simulated year on NCAR's Derecho supercomputer) despite no optimization effort for CAM-NN and having retained some of the ZM scheme to compute CAPE and convective mass flux. Future implementations could be optimized for greater computational efficiency.

\subsection{Observational and reanalysis datasets}\label{subsec:obs}

We use multiple observational and reanalysis datasets to evaluate model performance. We use the same time period for these datasets as for the CAM6 simulations analysis period (01/2001–12/2014).

For precipitation we use the GPCP v3.2 monthly dataset (gpcp\_v3.2\_monthly) which provides globally gridded, monthly-mean precipitation estimates derived from a combination of satellite observations and in situ rain gauge data \cite{huffman2023new}. We also use 30-mn precipitation from the Integrated Multi-satellite Retrievals for GPM (IMERG), which merges passive microwave, infrared, and rain gauge, resulting in a dataset with high temporal and spatial resolution (Huffman et al., 2020). IMERG is particularly successful over oceans, where rain gauge data are sparse or unavailable \cite{pradhan2023performance}.

For analysis of tropical convectively coupled waves, we use daily outgoing longwave radiation (OLR) provided by the NOAA Climate Data Record (CDR) Version 2.0 \cite{lee2025noaa}. This OLR data is available in global $1^{\circ}$ equal-angle gridded format, spanning January 1, 1979 to present. The OLR CDR products combine high-resolution infrared radiation sounder (HIRS) radiance observations and infrared hyperspectral radiance data from the Infrared Atmospheric Sounding Interferometer (IASI) and the Cross-track Infrared Sounder (CrIS) instruments.

We evaluate net radiative fluxes at the top of the atmosphere (TOA) using monthly data from the Clouds and the Earth’s Radiant Energy System (CERES) project, which provides measurements of Earth’s radiation budget acquired by broadband radiometers onboard multiple NASA satellite platforms \cite{loeb2018clouds}. These instruments measure outgoing longwave and reflected shortwave radiation, enabling accurate estimates of radiative fluxes and energy balance at the TOA.

Cloud cover fraction was obtained from the CALIPSO-lidar-only monthly product of the 3S-GEOPROF-COMB dataset \cite{bertrand2024global}, gridded globally at 2.5$^{\circ}$ $\times$ 2.5$^{\circ}$ resolution. We used the period 1 January 2006 to 31 December 2014 to overlap with our simulations.

ERA5 is the fifth-generation global atmospheric reanalysis dataset produced by the European Centre for Medium-Range Weather Forecasts (ECMWF; \citeNP{hersbach2020era5}). We use ERA5 as a reference dataset to evaluate atmospheric temperature and relative humidity simulated by CAM6, assessing how well the various configurations we test (i.e. CAM-NN, CAM-ZM) captures the observed climatological means.

\section{Results}\label{sec:results}

\subsection{Climatological means}\label{subsec:climato}

Both CAM6 versions tested (CAM-NN and CAM-ZM) run stably for the full 15 years period of this study (01/01/2000 - 12/31/2014) which includes the spinup period. NN-CONV conserves total moisture (precipitating and non-precipitating) when implemented in CAM6, as confirmed by NN-CONV satisfying CAM6's water and energy checkers, and consistent with the conservation properties built into its architecture.

\begin{figure}[H]
\centering
\includegraphics[width=\textwidth]{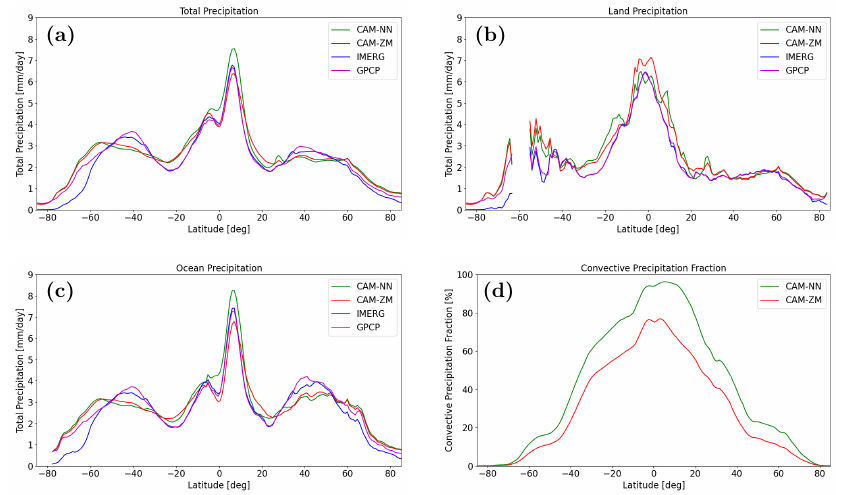}
\caption{Annual and zonal-mean precipitation (mm/day) for CAM-ZM, CAM-NN, IMERG, and GPCP (2001--2014), shown for (a) all regions, (b) land only, and (c) ocean only. Total precipitation is shown including convective and large-scale precipitation. Panel (d) shows the zonal-mean convective precipitation fraction for CAM-ZM and CAM-NN, computed at each grid point as the ratio of monthly-mean convective to total precipitation and then averaged annually and zonally.}
\label{fig:Fig3}
\end{figure}

\begin{figure}[H]
\centering
\includegraphics[width=\textwidth]{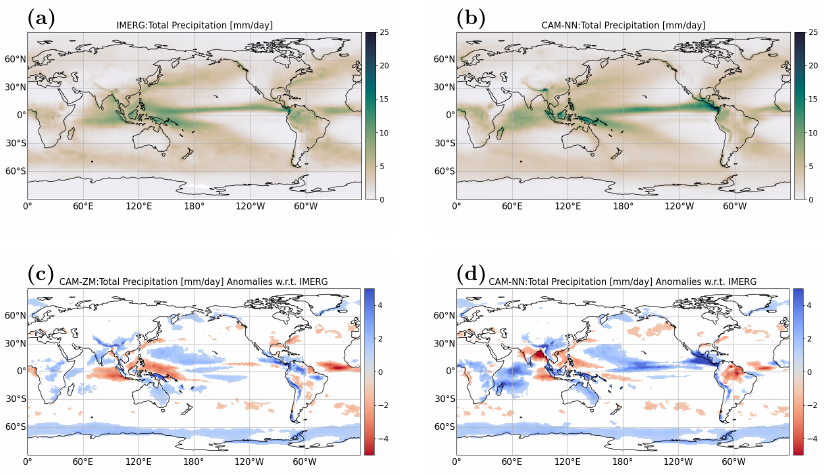}
\caption{Annual-mean precipitation climatology (mm/day) for (a) IMERG observations and (b) CAM-NN. Panels (c) and (d) show the bias relative to IMERG for CAM-ZM and CAM-NN, respectively. Regions with biases within $\pm$2 mm/day are masked. Global area-weighted root mean square error (RMSE) is 1.00 mm/day for CAM-ZM (c) versus 1.27 mm/day for CAM-ZM (d).} 
\label{fig:Fig4}
\end{figure}

When compared to observations and CAM-ZM over 2001–2014, CAM-NN captures key features of the meridional variability of the zonal-mean precipitation (Fig.~\ref{fig:Fig3}a), including the sharp maxima in the Intertropical Convergence Zone (ITCZ), the extent and intensity of subtropical dry zones, and the wet maxima in the extratropical storm tracks. CAM-NN overestimates precipitation rates in the deep tropics in the zonal mean by about 10\%. This wet bias for CAM-NN is found over the ocean and absent over land, whereas CAM-ZM has a compensating wet bias over land and dry bias over ocean (Fig.~\ref{fig:Fig3}b,c). At middle and high latitudes, deep convection is less important than in the tropics and CAM-NN and CAM-ZM have relatively similar mean precipitation to each other as compared to observations. Convective precipitation represents a higher fraction of total rainfall in CAM-NN than in CAM-ZM at all latitudes (Fig.~\ref{fig:Fig3}d). At regional scales, CAM-NN (Fig.~\ref{fig:Fig4}b) reproduces the observed (Fig.~\ref{fig:Fig4}a) annual-mean precipitation pattern reasonably well, including the Pacific and Atlantic ITCZ, the maximum over the Maritime Continent, and secondary maxima over monsoonal regions such as the Amazon, equatorial Africa, and India. Climatological biases are larger in CAM-NN than in CAM-ZM (Fig.~\ref{fig:Fig4}c,d; global area-weighted root mean square error (RMSE) is 1.27 mm/day for CAM-NN versus 1.00 mm/day for CAM-ZM), though not dramatically so given that the remaining model parameterizations have not been retuned to account for the new convection scheme, unlike default CAM6 where the full suite of parameterizations has been collectively calibrated. Both models share a similar bias pattern over the tropical Pacific, with precipitation underestimated over the West Pacific and overestimated over the East Pacific and tropical Atlantic. In CAM-NN, the positive bias over the west coast of Central America is particularly pronounced. An additional positive bias appears in CAM-NN over the Indian Ocean just north of Madagascar. Despite these regional biases, CAM-NN produces a broadly realistic climatology, and its precipitation biases are not substantially worse than those of CAM-ZM.

The vertical and latitudinal structure of relative humidity (Fig.~S2 in Supporting Information S1) and cloud fraction (Fig.~S3 in Supporting Information S1) in CAM-NN show reasonable agreement with reanalysis and observations, respectively. Biases are slightly worse in CAM-NN compared to CAM-ZM, likely due in part to a lack of proper calibration with CAM-NN. The sharp cutoff of the NN-CONV scheme at a specific sigma level near the tropopause is also evident in relative humidity. Temperature anomalies with respect to reanalysis are comparable between CAM-NN and CAM-ZM (Fig.~S4 in Supporting Information S1). The top-of-atmosphere (TOA) energy balance likewise shows radiative flux anomalies with respect to observations of similar magnitude between CAM-ZM and CAM-NN (Fig.~S5 in Supporting Information S1).

\subsection{Precipitation intensity}\label{subsec:intensity}

\begin{figure}[H]
\centering
\includegraphics[width=0.9\textwidth]{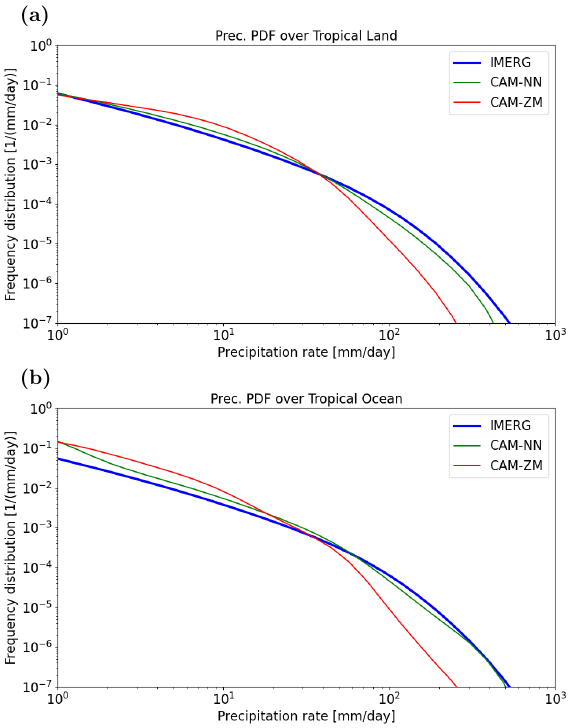}
\caption{Precipitation intensity distribution for IMERG observations (blue), CAM-NN (green) and CAM-ZM (red) from 6-hourly data over (a) tropical land and (b) tropical oceans. Histograms are computed on a logarithmic scale (30 bins, 1--1000 mm/day), and then normalized by total sample size and bin width to yield a probability density function [1/(mm/day)].}
\label{fig:Fig5}
\end{figure}

Going beyond the mean climatology, two key hydrological deficiencies of conventional convection schemes are a drizzle bias, i.e. an overproduction of weak precipitation events (Stephens et al., 2010) and large underestimates or overestimates of precipitation extremes depending on the climate model \cite{wehner2020characterization}. Performance on precipitation extremes is particularly non-robust in the tropics \cite{ogorman2012sensitivity}. To assess these issues, we compute a probability density function (PDF) by binning 6-hourly precipitation intensity on a logarithmic scale (30 bins, 1–1000 mm/day) and normalizing by total sample size and bin width. We compare this PDF for CAM-ZM and CAM-NN against 6-hourly IMERG observations conservatively regridded to the CAM6 grid, separately for tropical land and ocean ($30^{\circ}\text{S--}30^{\circ}\text{N}$; Fig.~\ref{fig:Fig5}). 

CAM-ZM exhibits the well-documented light precipitation bias, with excessive weak precipitation events, and also has a severe underrepresentation of extreme events (Fig.~\ref{fig:Fig5}a,b). CAM-NN substantially reduces these biases, producing a more realistic distribution over both land and ocean, with particularly strong agreement over land (Fig.~\ref{fig:Fig5}a,b). This large improvement in precipitation intensity distribution may be expected to be highly beneficial both for impacts modeling (e.g., allowing for a much easier task for statistical downscaling and bias correction of precipitation) and for coupling to model representations of land hydrology and vegetation which are sensitive to precipitation intensity.

\subsection{Diurnal cycle of precipitation}\label{subsec:diurnal}

Another key metric of sub-seasonal variability is the diurnal cycle of precipitation, which is also poorly represented in most global climate models. We compute a mean diurnal composite from hourly CAM6 output by averaging, for each location and hour-of-day bin, the precipitation over all available days during local summer (July-August in the Northern Hemisphere, December-February in the Southern Hemisphere), and then convert UTC timestamps to local solar time (LST). An area-averaged diurnal cycle over tropical land ($30^{\circ}\text{S--}30^{\circ}\text{N}$) is obtained by weighting each grid point by its surface area and land-area mask, then averaging spatially. The peak precipitation hour at each grid point is identified as the LST of maximum mean precipitation, with grid points showing weak diurnal amplitude masked out. This diagnostic is applied to both CAM6 simulations (CAM-ZM and CAM-NN) as well as to hourly IMERG observations conservatively regridded to the CAM6 grid, enabling a consistent comparison of the simulated and observed diurnal precipitation cycle. 

\begin{figure}[H]
\centering
\includegraphics[width=0.9\textwidth]{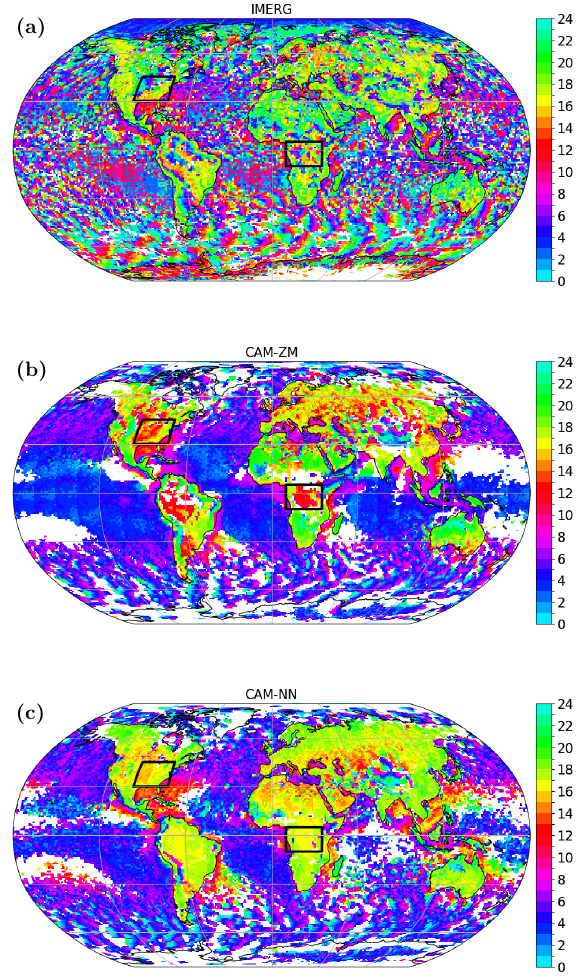}
\caption{Local solar time of diurnal precipitation peak averaged over local summer (July--August in the Northern Hemisphere, January--February in the Southern Hemisphere) for (a) IMERG observations, (b) CAM with the ZM convection scheme (CAM-ZM), and (c) CAM with the neural network convection scheme (CAM-NN). Areas where the diurnal amplitude is small (ratio of first harmonic diurnal amplitude to mean precipitation lower than 0.1) are masked. The black boxes indicate the regions used to define the regional-mean diurnal cycles shown in Fig.~\ref{fig:Fig7}.}
\label{fig:Fig6}
\end{figure}

\begin{figure}[H]
\centering
\includegraphics[width=\textwidth]{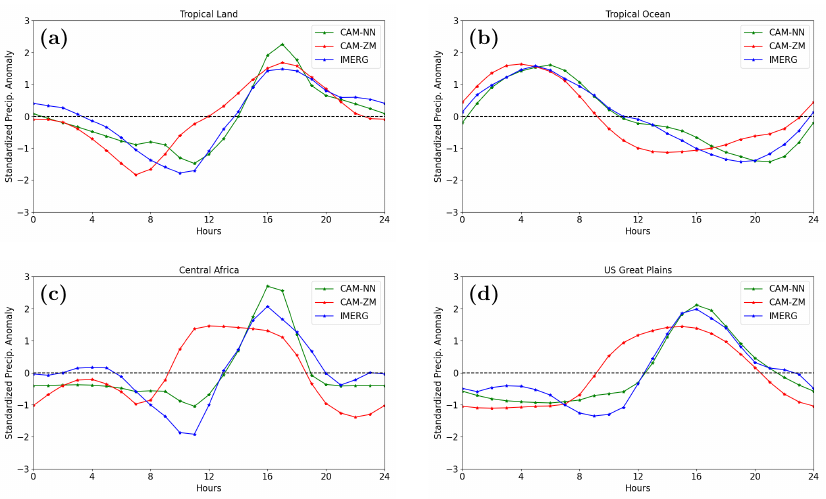}
\caption{Diurnal cycle of precipitation anomaly (computed as the deviation from the diurnal-mean value, normalized by the standard deviation of the diurnal cycle) versus local solar time averaged over (a) tropical land (30$^{\circ}$S--30$^{\circ}$N, land only), (b) tropical oceans (30$^{\circ}$S--30$^{\circ}$N, ocean only), (c) Central Africa and (d) the Great Plains and Eastern US (see region outlines on Fig.~\ref{fig:Fig6}), during local summer (July--August in the Northern Hemisphere, January--February in the Southern Hemisphere). Lines show CAM-NN (green), CAM-ZM (red), and IMERG observations (blue). 
Root mean square error (RMSE) is 0.37 for CAM-NN vs. 0.63 for CAM-ZM over tropical lands (a), 0.21 for CAM-NN vs. 0.50 for CAM-ZM over tropical oceans (b), 0.31 for CAM-NN vs. 0.85 for CAM-ZM over Great Plains and Eastern US (c), and 0.50 for CAM-NN vs. 1.24 CAM-ZM over Central Africa (d).}
\label{fig:Fig7}
\end{figure}

Observations indicate that precipitation over land generally peaks in the late afternoon to early evening, while over the ocean it peaks during the early morning hours (Fig.~\ref{fig:Fig6}a). The diurnal cycle of precipitation is primarily controlled by the deep convection scheme, and CAM-ZM produces a premature midday precipitation peak over land, most notably over South America, the North American Great Plains, Central Africa, and large parts of Siberia, as well as an overly homogeneous peak hour over ocean that lacks the spatial variability seen in observations (Fig.~\ref{fig:Fig6}b). CAM-NN, on the other hand, captures a more realistic spatial distribution of peak precipitation timing, with precipitation peaking around 17:00 LST over South America, Central Africa, and most of Eurasia (Fig.~\ref{fig:Fig6}c), in good agreement with observations. In addition there is more realistic spatial variability in peak timing over tropical ocean. These are notable improvements, particularly given the relatively coarse model resolution (\~100 km); models that do not resolve convection struggle to accurately simulate the diurnal cycle even at much higher spatial resolution \cite{christopoulos2021assessing,dirmeyer2012simulating}. When averaged spatially over all tropical land and ocean regions, the composite diurnal cycle confirms that CAM-NN closely tracks the timing of the observed late-afternoon peak and mid-morning minimum over land (Fig.~\ref{fig:Fig7}a) and reproduces the timing of the nighttime maximum and daytime minimum over oceans (Fig.~\ref{fig:Fig7}b). When compared to observations, the root mean square error (RMSE) for the shape of the precipitation diurnal cycle is 0.37 for CAM-NN vs. 0.63 for CAM-ZM over tropical land, and 0.21 vs. 0.50 for CAM-NN and CAM-ZM over tropical oceans, demonstrating substantial quantitative gains over land and ocean. While limitations in the simulated diurnal cycle remain (e.g. underestimation of nighttime rainfall activity over land compared to observations, as seen by nighttime peak hours in IMERG over tropical land -- e.g. the Amazon or Central Africa -- being absent in either CAM-NN or CAM-ZM; see Fig.~\ref{fig:Fig6}), CAM-NN markedly improves the peak precipitation timing compared to CAM-ZM in both tropical land regions such as Central Africa (Fig.~\ref{fig:Fig7}d) and midlatitude land regions such as the Great Plains and Eastern US (Fig.~\ref{fig:Fig7}c), during local summer.

\subsection{Tropical wave activity}\label{subsec:waves}

\begin{figure}[H]
\centering
\includegraphics[width=\textwidth]{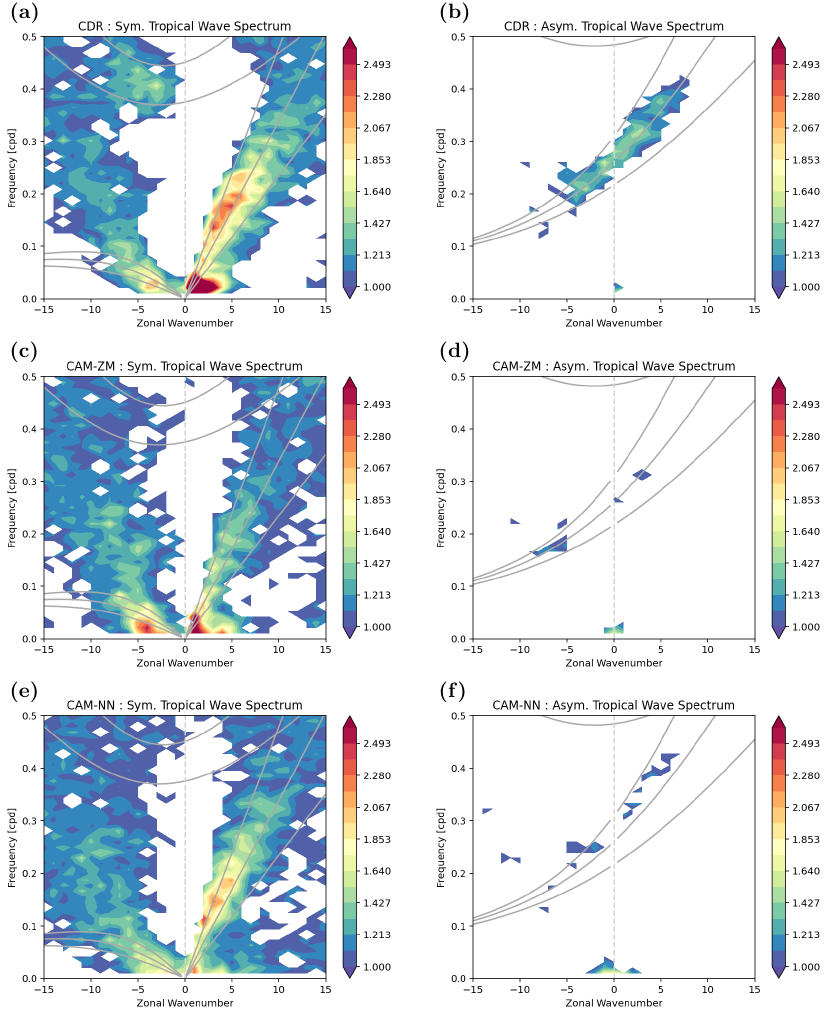}
\caption{Wheeler--Kiladis wavenumber-frequency power spectra of OLR, normalized by the background spectrum, for the (left) symmetric and (right) antisymmetric components averaged over the tropical belt (10$^{\circ}$S--10$^{\circ}$N). Panels show (a),(b) observations from NOAA's Climate Data Record (CDR), (c),(d) CAM-ZM, and (e),(f) CAM-NN. Shallow water dispersion curves are overlaid. Values above 1 indicate power above the background, while values below 1 are shown as blank.}
\label{fig:Fig8}
\end{figure}

Finally, we analyze convectively-coupled wave activity using the Wheeler-Kiladis space-time spectral decomposition \cite{wheeler1999convectively}. Daily-mean outgoing longwave radiation (OLR) over the tropical belt $10^{\circ}\text{S--}10^{\circ}\text{N}$ is decomposed into symmetric and antisymmetric components about the equator, and analyzed spectrally over overlapping 96-day segments with 65-day overlap. The raw power spectra are normalized by a smoothed background spectrum to identify statistically significant wave signals relative to a red-noise background (Fig.~\ref{fig:Fig8}). Global climate models typically underestimate tropical wave activity, and CAM-ZM is no exception: it shows weak Kelvin and Yanai (mixed Rossby-gravity) wave signals and lacks a clear MJO signal (Fig.~\ref{fig:Fig8}c,d) relative to observations (Fig.~\ref{fig:Fig8}a,b). CAM-NN exhibits enhanced Kelvin wave activity in better agreement with observations, and the Yanai wave signal shows modest improvement (Fig.~\ref{fig:Fig8}e,f). The MJO signal in CAM-NN, however, is much weaker than observed (Fig.~\ref{fig:Fig8}a,e). This may reflect the fact that the MJO is not solely controlled by the deep convection scheme: surface flux feedbacks \cite{maloney2004surface,sobel2010surface} and other dynamical processes such as shallow convection \cite{cai2013impacts} are also known to play an important role. This suggests that improvements in the MJO may require retuning or further development of other parameterizations beyond deep convection.

\section{Conclusions}\label{sec:conclusions}

A key challenge in climate modeling is capturing how deep convection shapes the large-scale circulation and hydrological cycle, given the substantial gap between the scales at which convection actually occurs and the coarser scales that global climate models can resolve. Hybrid machine-learning approaches offer the potential to take advantage of improvements in accuracy from storm-resolving simulations but in climate models that retain physics for resolved components at coarse-resolution and thus can be run for much longer. Here, a deep neural network was used to represent deep moist convection, trained on high-resolution (2-4 km) gSAM simulation output. When NN-CONV replaces the traditional deep convection parameterization in CAM6, the resulting model produces stable 15-year simulations that conserve energy and moisture, consistent with the conservation properties built into the scheme's architecture.

CAM-NN captures the key features of the observed zonal-mean precipitation climatology, including the ITCZ, subtropical dry zones, and extratropical storm tracks. Regional biases in CAM-NN are somewhat larger than in CAM-ZM, likely because the remaining CAM6 parameterizations were not re-tuned to account for the new convection scheme, unlike the default configuration where the full suite of parameterizations has been collectively calibrated.

A key result is the substantial improvement in the intensity distribution of tropical precipitation. CAM-NN greatly reduces the well-documented light rainfall bias of CAM-ZM, the overproduction of weak precipitation events, while simultaneously improving the representation of intense extremes, bringing the distribution into closer agreement with IMERG observations over both land and ocean, with particularly strong agreement over land. Such improvements in precipitation intensity are expected to be beneficial for coupling to the land hydrology and vegetation and should also make it easier to statistically downscale precipitation for use in climate-change impacts research. This result is encouraging because coarse global climate models often simulate tropical precipitation extremes inconsistently, limiting their usefulness for predicting associated climate risks.

CAM-NN also produces a markedly more realistic diurnal cycle of precipitation. Over tropical land, it captures a late-afternoon to early-evening peak broadly consistent with observations, in contrast to CAM-ZM's premature midday peak in many regions. This improvement is especially notable given the coarse model resolution ($\sim 100 \text{ km}$); accurate simulation of the diurnal cycle remains challenging even for higher-resolution models that do not resolve convection. Thus the use of a neural network deep-convection parameterization is able to bring some of the improvements from convection-resolving models to much coarser resolutions for use in climate simulations.

The picture is more mixed for tropical wave activity. CAM-NN enhances Kelvin wave power and shows modest improvement in the Yanai wave signal relative to CAM-ZM, bringing both into closer agreement with observations. The MJO signal, however, remains weaker than observed. This residual bias may reflect limitations in NN-CONV itself, but is also likely attributable to processes outside the deep convection scheme, including surface flux feedbacks and shallow convection, suggesting that further improvement will require retuning or further development of other parameterizations in addition to convection.

An important practical advantage of the NN-CONV approach is that it retains adjustable parameters that provide tuning knobs analogous to those available in conventional schemes (e.g., cloud ice mass-to-number conversion factor, minimum value for CAPE to activate the convection scheme). In this study, only the cloud ice mass to number conversion factor was retuned, but the broader CAM6 parameter space was left unchanged. A systematic retuning of the full model configuration could be expected to reduce the remaining climatological biases and further improve performance.

Our results demonstrate an ability to transfer NN-CONV from its training model (gSAM) to a structurally different target model (CAM6), with a different vertical grid and set of prognostic variables. The successful transfer suggests that retraining is not necessary when porting the scheme to a new vertical grid (simple interpolation of the outputs appears sufficient), broadening the potential applicability of the approach to other climate models. In addition, we find that training in boreal winter is sufficient to obtain good performance in other seasons, such as for the diurnal cycle in boreal summer in the Northern Hemisphere.

Our results also demonstrate the feasibility of using a neural network to replace just one aspect of moist physics (here the deep convection parameterization) without also replacing the grid-scale cloud macrophysics and microphysics, boundary layer turbulence, or radiation schemes. Importantly, the ability to avoid replacing the radiation scheme should make the resulting climate model more versatile and able to generalize under climate forcings. For climate change, it would still be necessary to train on warmer or colder climates to avoid distribution shifts of the inputs \cite{ogorman2018using,rasp2018deep}, or to possibly avoid these shifts using more climate-invariant inputs that are found through knowledge of physics \cite{beucler2024climate} or learned from data \cite{liu2026cera}. However, because NN-CONV doesn't involve radiation and only indirectly affects cloud radiative properties, we do not expect that training on additional expensive high-resolutions simulations would be needed for the fast response to CO$_2$ forcing or for the response to different ozone distributions or aerosols emissions, and this could be investigated in future research.

Future research could also explore the addition of neural-network predictions for convective momentum tendencies and the influence of convection on surface fluxes through changes in gustiness, as well as the development of non-local parameterizations that account for convective interactions across neighboring grid columns. Additional insights on the generalizability of this framework may be gained by applying NN-CONV to coupled atmosphere-ocean CESM2 simulations, to different resolutions of CAM6 (e.g., at 25km horizontal resolution), to different climate scenarios, or to other earth system models.

% FIGURES

% TABLES 
\begin{table}[H]
\caption{Summary of the specific hyperparameters and neural network architecture that were found to lead to good offline performance on the test dataset and online performance in CAM6 and are used for the results presented in the paper.} \label{tab:Table1}
\centering
\begin{tabular}{l l}
    \hline
        Training hyperparameters  & Configuration used \\
   \hline
      Training Data Volume  & 34 days of global hourly snapshots   \\
      Loss Function  & Huber loss   \\
      Learning Rate  & 0.002   \\
      Size of Hidden Layers & 128 nodes x 5 layers \\
      Optimizer & Adam \\
      Activation Function & ReLU \\
      Epochs  & 55 \\
    \hline
\end{tabular}
\end{table}

\section*{Acknowledgments}

This project was supported by Schmidt Sciences. Xavier Levine, Judith Berner, Paul O’Gorman, and Griffin Mooers were supported through the M2LInES project. Jack Atkinson and Dominic Orchard were supported through ICCS (the Institute of Computing for Climate Science). Marat Khairoutdinov was supported by the NSF Grant AGS-2218827 to Stony Brook University. We thank William Chapman for valuable discussions on implementing NN-CONV in CAM6 and for his guidance with running and analyzing simulations. We thank Marion Weinzierl for her software support in setting up timing and benchmarking in CAM6. We acknowledge insightful discussions with Hugh Morrison, Richard Neale, Brian Medeiros, and Vincent Larson about convective mixing in CAM6. We thank Ignacio Lopez for his helpful comments. Training of the parameterization was performed on Bridges2 at the Pittsburgh Supercomputing Center through allocation EES240105 from the Advanced Cyberinfrastructure Coordination Ecosystem: Services \& Support (ACCESS; \citeNP{boerner2023access}) program, which is supported by U.S. National Science Foundation grants \#2138259, \#2138286, \#2138307, \#2137603, and \#2138296. CAM6 simulations and analysis were done on NCAR’s Derecho supercomputer, and we gratefully acknowledge the support of the NCAR team. The authors further acknowledge the MIT Office of Research Computing and Data for providing high performance computing resources through MIT ENGAGING that have contributed to the research results reported within this paper. Additionally, this work was supported by the U.S. National Science Foundation National Center for Atmospheric Research (NSF NCAR), which is a major facility sponsored by the NSF under Cooperative Agreement 1852977. Any opinions, findings, and conclusions or recommendations expressed in this material are those of the author(s) and do not necessarily reflect the views of the U.S. National Science Foundation.

% Your Data Availability Statement is an unnumbered section right before the bibliography.
\section*{Open Research Statement}

The source code for CAM6 used in this study, including both the standard configuration and the machine-learning-enabled version incorporating the neural network convection scheme, is archived on GitHub at \url{https://github.com/m2lines/CAM-ML}. The Python scripts used to produce the analysis and figures in this manuscript are archived on GitHub at \url{https://github.com/xlevine/CAMML_JAMES}. The code used to train the neural network convection parameterization is archived on Zenodo at \url{https://doi.org/10.5281/zenodo.22122596}. The gSAM simulation data used for training were produced as part of the DYAMOND2 project \cite{stevens2019dyamond}; the experimental protocol is publicly available at \url{https://easy.gems.dkrz.de/DYAMOND/index.html}.

\section*{Conflicts of Interest}

The authors declare no conflicts of interest relevant to this study.

% Bibliography
%\cite{*} % If you have uncited bibliography items.
%
\bibliography{CAMML} % Replace ``wiley'' with your bibliography's file name.  Do not specify its extension, e.g., .bib.

\clearpage
\begin{center}\includegraphics[page=1,width=\textwidth,height=0.95\textheight,keepaspectratio]{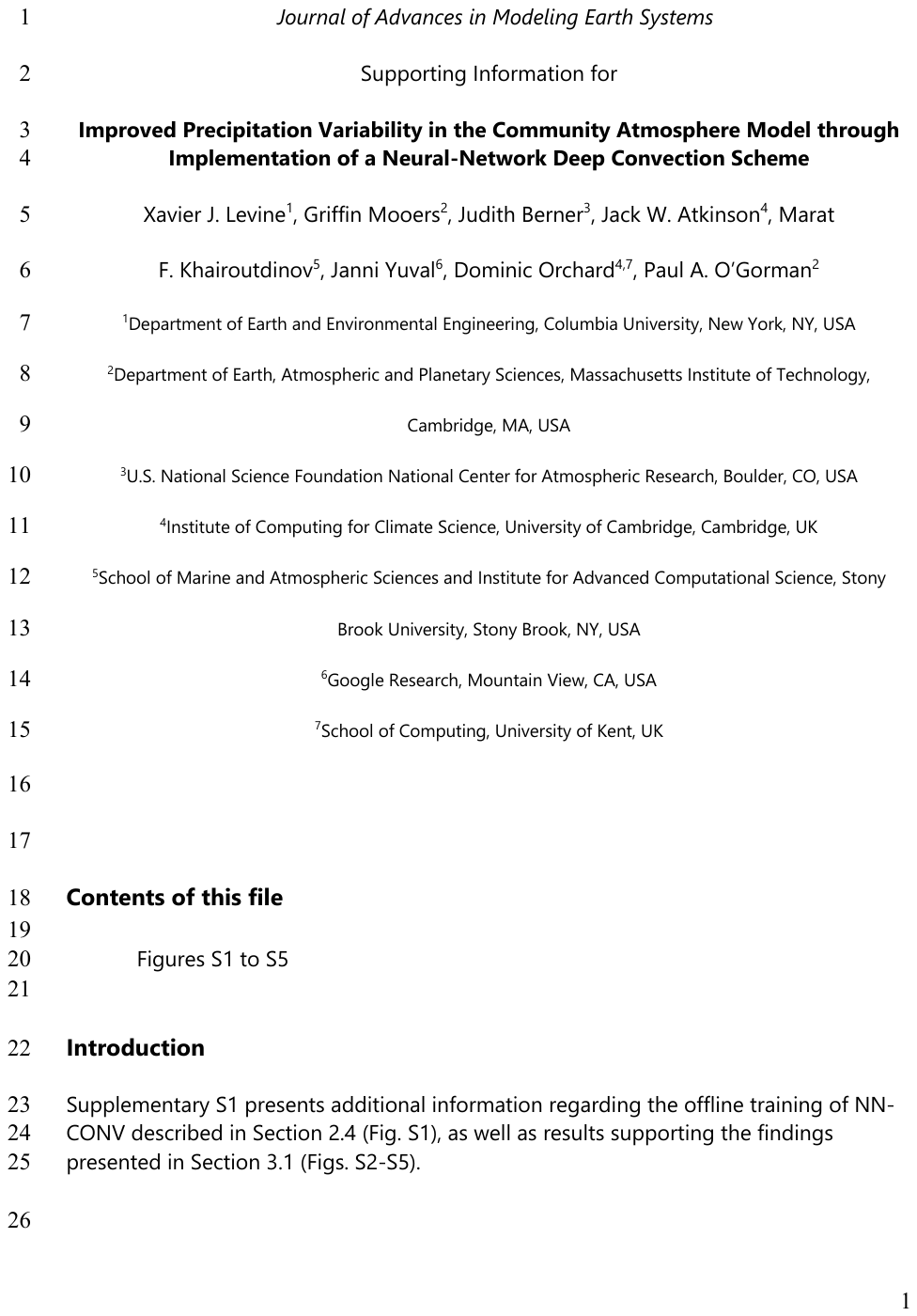}\end{center}
\clearpage
\begin{center}\includegraphics[page=2,width=\textwidth,height=0.95\textheight,keepaspectratio]{LevineAL_Supplementary.pdf}\end{center}
\clearpage
\begin{center}\includegraphics[page=3,width=\textwidth,height=0.95\textheight,keepaspectratio]{LevineAL_Supplementary.pdf}\end{center}
\clearpage
\begin{center}\includegraphics[page=4,width=\textwidth,height=0.95\textheight,keepaspectratio]{LevineAL_Supplementary.pdf}\end{center}
\clearpage
\begin{center}\includegraphics[page=5,width=\textwidth,height=0.95\textheight,keepaspectratio]{LevineAL_Supplementary.pdf}\end{center}
\clearpage

\end{document}